\documentclass[twocolumn,superscriptaddress,preprintnumbers,amsmath,amssymb,prl]{revtex4}
\usepackage{graphicx}
\usepackage{dcolumn,xcolor,ulem}
\usepackage{amsmath} 
\usepackage{amssymb}
\usepackage{amsfonts}
\usepackage{bm}
\usepackage[latin1]{inputenc}

\begin{document}

\title{Multiple yielding processes in a colloidal gel under large amplitude oscillatory stress}

\author{Thomas~Gibaud}
\email{thomas.gibaud@ens-lyon.fr}
\affiliation{Universit\'e de Lyon, Laboratoire de Physique, \'Ecole Normale Sup\'erieure de Lyon, CNRS UMR 5672, 46 All\'ee d'Italie, 69364 Lyon cedex 07, France}
\author{Christophe~Perge}
\affiliation{Universit\'e de Lyon, Laboratoire de Physique, \'Ecole Normale Sup\'erieure de Lyon, CNRS UMR 5672, 46 All\'ee d'Italie, 69364 Lyon cedex 07, France}
\author{Stefan~B.~Lindstr{\"o}m}
\affiliation{Solid Mechanics, Department of Management and Engineering, Link\"oping University, SE-581 83 Link\"oping, Sweden}
\author{Nicolas~Taberlet}
\affiliation{Universit\'e de Lyon, Laboratoire de Physique, \'Ecole Normale Sup\'erieure de Lyon, CNRS UMR 5672, 46 All\'ee d'Italie, 69364 Lyon cedex 07, France}
\affiliation{Universit\'e de Lyon, UFR de Physique, Universit\'e Claude Bernard Lyon I, Lyon, France}
\author{S{\'e}bastien~Manneville}
\affiliation{Universit\'e de Lyon, Laboratoire de Physique, \'Ecole Normale Sup\'erieure de Lyon, CNRS UMR 5672, 46 All\'ee d'Italie, 69364 Lyon cedex 07, France}
\affiliation{Institut Universitaire de France}

\date{\today}

\begin{abstract}
Fatigue refers to the changes in material properties caused by repeatedly applied loads. It has been widely studied for, e.g., construction materials, but much less has been done on soft materials. Here, we characterize the fatigue dynamics of a colloidal gel. Fatigue is induced by large amplitude oscillatory stress (LAOStress), and the local displacements of the gel are measured through high-frequency ultrasonic imaging. We show that fatigue eventually leads to rupture and fluidization. We evidence four successive steps associated with these dynamics: (i)~the gel first remains solid, (ii)~it then slides against the walls, (iii)~the bulk of the sample becomes  heterogeneous and displays solid--fluid coexistence, and (iv)~it is finally fully fluidized. It is possible to homogeneously scale the duration of each step with respect to the stress oscillation amplitude $\sigma_0$. The data are compatible with both exponential and power-law scalings with $\sigma_0$, which hints at two possible interpretations in terms of delayed yielding in terms activated processes or of the Basquin law. Surprisingly, we find that the model parameters behave nonmonotonically as we change the oscillation frequency and/or the gel concentration.
\end{abstract}
\maketitle

\section{INTRODUCTION}%-------------------------------------------------------------------------
Fatigue refers to the changes in material properties caused by repeatedly applied loads \cite{Suresh}. Fatigue is quite universal and affects a broad range of materials from metals \cite{Bhat} to biomaterials such as adhesion clusters of cells \cite{Erdmann}. In metals, the microscopic events associated with fatigue depend on the load and usually weaken the material. For example, if loads are above a certain threshold, microscopic cracks form at stress concentrators such as the surface, persistent slip bands, and grain boundaries \cite{Suresh,Thompson, Frost, Murakami}. Eventually a crack reaches a critical size, propagates, and the structure fractures \cite{Griffith, Paris}. Here, we address the question of fatigue in a soft material namely a colloidal gel. Under a constant stress, colloidal gels, like many other viscoelastic solids such as foams or emulsions, flow only above a critical stress referred to as the yield stress $\sigma_y$. This stress-induced fluidization is involved in virtually any application of colloidal gels, and it is essential to correctly measure and control this yield stress to ensure the safe and proper use of the material.
Yet it has been shown in recent years that many phenomena may affect the way a material yields \cite{Barnes, Nguyen, Moller,Bonn}. Slippery boundary conditions may induce the solid-body motion of the material without bulk fluidization \cite{Bonn}. Long and complex transient regimes may show coexistence of fluid and solid phases before a liquid-like stationary state is reached \cite{Divoux}. In particular colloidal gels are prone to so-called ``delayed yielding'' where it may take a very long time for a weak gel to yield under a constant stress \cite{Gopalakrishnan,Sprakel, Lindstrom}. For instance, a gel can fail under its own weight after several hours or days \cite{Allain,Starrs}. This ``gravitational collapse'' has been associated to the progressive weakening of the gel network due to thermally activated localized rearrangements \cite{Bartlett} or to the interplay between the poroelasticity of the gel and gravitational compression \cite{Buzzaccaro, Manley}. Delayed yielding has also been investigated in gels subjected to a constant shear stress imposed by a rheometer, including carbon black gels \cite{Gibaud,Grenard}, thermo-reversible protein gels \cite{Brenner} or weak gels of polystyrene particles \cite{Sprakel}. In all cases, the gels are eventually fluidized after complex transient regimes. These dynamics have been related to activated processes: the applied shear stress $\sigma$ decreases the energy barrier for bond breakage leading to failure times that decrease exponentially with $\sigma$ \cite{Gopalakrishnan, Lindstrom}. In this work we focus on the interplay between fatigue and yielding of carbon black gels under cyclic shear, a phenomenon which is comparatively much less documented in soft matter than the case of constant load.

In a previous paper devoted to large amplitude oscillatory shear stress (LAOStress) experiments \cite{Perge}, we have first shown that the classical procedure to determine the yield stress $\sigma_y$ displays surprising results. In an oscillatory experiment at a constant frequency $f$, where the amplitude $\sigma_0$ of the shear stress is ramped up with time $t$, $\sigma_y$ is defined as the stress amplitude at which the loss modulus $G''$ overcomes the elastic modulus $G'$~\cite{Macosko, Larson,Russel}. In carbon black gels, $\sigma_y$ depends on the velocity of the applied stress ramp \cite{Perge}: $\sigma_y$ decreases as the ramp rate increases. This result is inconsistent with physical aging, where both $G'$ and $\sigma_y$ have been shown to increase\cite{negi}. This rate dependence of $\sigma_y$ is here one of the hallmarks of fatigue. To better characterize this fatigue-induced yielding, we performed further LAOStress experiments\cite{Perge}, this time keeping  the stress amplitude constant to $\sigma_0=11$~Pa. An extensive time-resolved analysis of the strain response, of the rheological Fourier spectra, and of standard Lissajous plots (stress vs strain or shear rate) coupled to simultaneous ultrasonic imaging, showed that the gel dynamics involved two different timescales, $\tau_w < \tau_f$, where $\tau_w$ is associated with failure at the walls, and $\tau_f$ with a slower heterogeneous fluidization involving solid--fluid coexistence until the whole sample becomes liquid-like. The spatial heterogeneities observed as the gel slowly fluidizes suggested a fragmentation of the initially solid-like gel into macroscopic domains eroded by the surrounding fluidized suspension.

Here, we thoroughly address the influence of the stress amplitude and of the frequency, as well as that of the carbon black concentration, on such a scenario. In Sect.~\ref{s.materials}, we first recall the sample preparation method and the specifications of the apparatus for combining standard LAOStress rheology and ultrasonic imaging in order to probe the local displacement within carbon black gels during LAOStress experiments. In Sect.~\ref{s.results}, we show that $\tau_f$ results from the accumulation of multiple yielding processes: the gel first remains solid, at $\tau_w$ it starts sliding against the walls, at $\tau_b$ it displays a bulk solid--fluid coexistence, and at $\tau_f$ it is finally fully fluidized. While the raw times $\tau_w$, $\tau_b$ and $\tau_f$ seem to display two different behaviours for low vs high stresses $\sigma_0$ as also observed under a constant load in Refs.~\cite{Gopalakrishnan,Sprakel, Lindstrom,Gibaud,Grenard}, scaling the durations of the successive yielding processes $\tau_w$, $\tau_b-\tau_w$ and $\tau_f-\tau_b$ with $\sigma_0$ allows us to describe the whole stress range within a single scaling. These results are further discussed in Sect.~\ref{s.discuss} where we compare our data with an extension of the delayed failure model \cite{Lindstrom} to oscillatory shear (Appendix 2) and with the Basquin law. The extension of the delayed failure model is based on activated processes and predicts that the characteristic durations decrease exponentially with $\sigma_0$, whereas the Basquin law predict a power-law behaviour.  Both models fit our data well given the small accessible range for $\sigma_0$.  Finally, we report a surprising behaviour of the parameters of the delayed yielding model and the Basquin law when varying the concentration of the gel and the frequency of the oscillations.

\section{MATERIALS AND METHODS}%-------------------------------------------------------------------------
\label{s.materials}
\subsection{Carbon black gel preparation}
Carbon black (CB) particles are colloidal, carbonated particles with a typical size range of 85 to 500~nm \cite{Trappe2000} that result from the partial combustion of hydrocarbon oils. These particles are widely used in the industry for mechanical reinforcement  or to enhance the electrical conductivity of plastic and rubber materials \cite{Donnet}. When dispersed in a mineral oil (density 0.838~g.cm$^{-3}$, viscosity 20 mPa.s, Sigma Aldrich), these CB particles (Cabot Vulcan XC72R of density 1.8~g.cm$^{-3}$) are weakly attractive with interactions of typical strength $U\sim30 k_B T$ \cite{Trappe2001,Trappe2007}, where $k_B$ is the Botzmann constant and $T$~is the absolute temperature. From a dispersed state, the particles aggregate to form sample-spanning networks of fractal dimension $d_f=2.2$ even at very low concentrations \cite{Trappe2000}. Here, we focus on such gel-forming dispersions at weight concentrations of 4, 6 and 8\%~w/w.

\subsection{Standard rheology experiments}
\label{s.rheol}

The mechanical properties of colloidal gels are typically measured using standard rheology experiments. Our rheological measurements are performed in a Taylor--Couette cell with smooth, polymethyl methacrylate (PMMA) walls (height 50 mm, rotating inner cylinder radius 48 mm, fixed outer cylinder radius 50 mm, gap width 2 mm) with a stress-controlled rheometer (ARG2, TA Instruments). The temperature is controlled by a water circulation around the Taylor--Couette cell and fixed to $25\pm0.1$ $^\circ$C for all experiments. To ensure a reproducible initial gel state, each measurement is prepared using the following sequence of steps: (i)~We preshear the suspension at 1000~s$^{-1}$ and then at -1000~s$^{-1}$ for 20~s each to break up any large aggregates. (ii)~We let the gel rest so that it can restructure by applying a zero shear rate for 2~s and then a zero shear stress for 10~s. (iii)~We probe the viscoelastic properties of the gel by monitoring the elastic modulus $G'$ and loss modulus $G''$ for 60~s at a very low stress in the linear regime ($\sigma_0=0.5$ Pa at $f=1$ Hz). (iv)~The sample is left to rest again for 10~s by applying a zero shear stress, which allows for possible residual stresses stored during the previous steps to relax \cite{Osuji,Grenard}. Finally, the LAOStress experiment is started by applying a sinusoidal stress of amplitude $\sigma_0$ and frequency $f$ starting at time $t=0$. The reader is referred to Refs.~\cite{Grenard,Perge} for more details, especially on internal stresses and on issues with inertia during LAOStress.

\subsection{Ultrasonic imaging under shear}

Several techniques allow for local flow characterization under shear, e.g. rheo-optics \cite{Pignon, Lerouge}, particle tracking \cite{Besseling1,Nordstrom,Jop}, magnetic resonance imaging \cite{Callaghan,Coussot} or ultrasonic velocimetry \cite{Manneville}. Here, we use high-frequency ultrasonic imaging \cite{Gallot}, a technique that is fast enough to follow the yielding dynamics while being insensitive to the opaqueness of the CB gel \cite{Perge}. To provide ultrasonic contrast the gel is seeded with 1\% w/w hollow glass spheres (Potters Sphericel, mean diameter 6~$\mu$m, mean density 1.1~g.cm$^{-3}$). This technique is implemented on the Taylor--Couette experiment described in Sect.~\ref{s.rheol}. A linear array of 128 piezoelectric transducers, immersed in the water tank and facing the outer cylinder of the Taylor--Couette device, sends short ultrasonic pulses with a central frequency of 15~MHz that propagate in a vertical plane inclined by about 10$^{\circ}$ from the normal to the outer cylinder. While traveling through the Taylor--Couette cell, these pulses get scattered by the hollow glass spheres suspended within the CB gel. For each incident plane pulse, the backscattered signal, corresponding to the interferences of the various echoes from the hollow glass spheres, is recorded by the transducer array, leading to an ``ultrasonic speckle'' signal with 128 measurement lines and typically 800 points sampled at 160~MHz. The data analysis consists in first processing the speckle signal into a beam-formed speckle image $S(r, z, t)$, where $r$ is the radial direction across the gap ($r=0$ being taken at the rotating inner cylinder and $r=2$~mm at the fixed outer cylinder) and $z$ is the vertical direction ($z=0$ corresponding to about 10~mm from the top of the Taylor--Couette device). Then two successive speckle images are cross-correlated in order to get the tangential displacement $\Delta$ as a function of $r$ and $z$. Here we set the time interval between two speckle images to five LAOStress oscillation periods. This allows us to map the displacement $\Delta(r, z, t)$ of the CB dispersion between successive states separated by five oscillation cycles over 32~mm along the vertical direction $z$ with a resolution of 250~$\mu$m and over the 2-mm gap with a resolution of 100~$\mu$m in the radial direction $r$. This temporal and spatial resolution is ideal for following the yielding dynamics of the CB dispersion under LAOStress without focusing on the intracycle dynamics \cite{Gibaud}. For full technical details, we refer the reader to Refs.~\cite{Perge,Gallot}.

%Graph1--------------------------------------------------------------------------
\begin{figure}
\centering
\includegraphics[width=7.5cm]{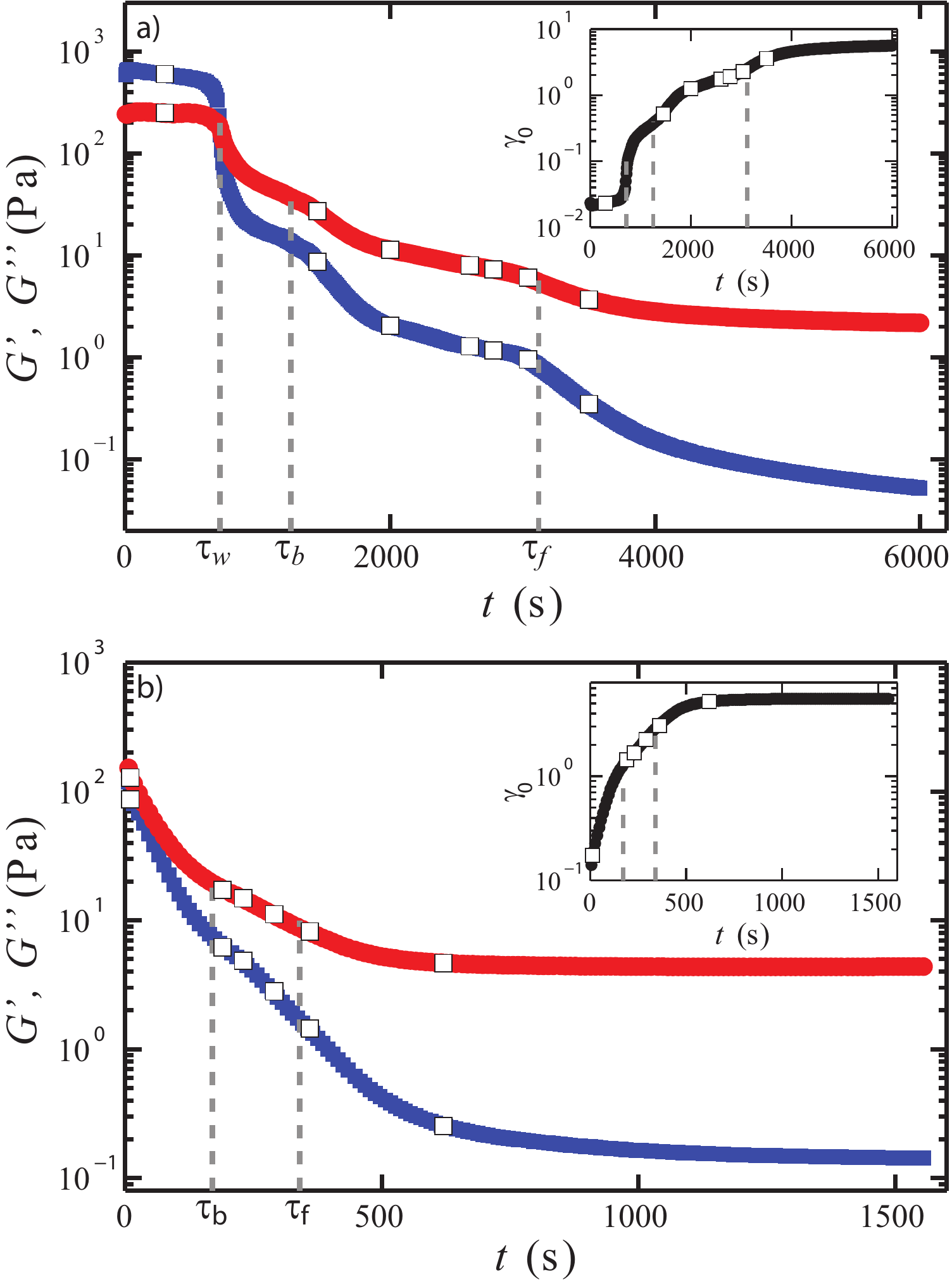}
\caption{Time evolution of the elastic modulus $G'$ (blue squares) and loss modulus $G''$ (red circles) for a carbon black dispersion at $c=6\%$~w/w under an oscillatory shear stress of constant amplitude $\sigma_0$ and frequency $f=1$~Hz. (a)~For a {\it low} stress amplitude, $\sigma_0= 15$~Pa, the gel initially displays a solid-like behaviour ($G'>G''$), then apparently yields ($G'=G''$) at $t\simeq 700$~s, and thereafter seems to flow like a liquid ($G'<G''$). (b)~For a {\it high} stress amplitude, $\sigma_0=27$~Pa, the gel seems to yield and flow like a liquid from the beginning of the experiment. The insets show the corresponding time evolutions of the strain amplitude $\gamma_0$. In the solid-like regime, the strain remains very low. $\gamma_0$ then increases by two orders of magnitudes and finally saturates at some maximum value in the fluid-like regime. White squares indicates the times at which the displacement field $\Delta(r, z, t)$, as obtained from  ultrasonic imaging, is illustrated in Fig.~\ref{figUSV} }
\label{figRheo}
\end{figure}

\subsection{Normal force measurements}
In order to measure the normal force $N$  during a LAOStress experiment, we switch from a Taylor--Couette geometry to a standard cone--plate  geometry with cone diameter 5~cm and angle 4$^{\circ}$. Like in the Taylor--Couette geometry, to ensure a reproducible initial state, each measurement is carried out by (i)~preshearing the suspension at 700~s$^{-1}$ then at -700~s$^{-1}$ for 60~s in each direction, (ii)~manually shearing the gel until normal forces cancel out, (iii)~letting the gel rest again for 30~s and (iv)~measuring $G'$ and $G''$ within the linear regime and checking that the normal force remains zero. Finally, we start the LAOStress experiment. Calibration experiments have shown that we can neglect the contribution of the centrifugal force to the measurements of $N$ provided that the shear rate is below 700~s$^{-1}$. We checked that the shear rate never exceeded 100~s$^{-1}$ during the LAOStress experiments.

\section{RESULTS}%-------------------------------------------------------------------------
\label{s.results}
\subsection{Standard rheological measurements}
Figure \ref{figRheo} reports the evolution of the elastic and loss moduli of carbon black dispersion at $c=6\%$ under LAOStress for both a ``low'' stress amplitude [$\sigma_0=15$~Pa, Fig.\ref{figRheo}(a)] and a ``high'' stress amplitude [$\sigma_0=27$~Pa, in Fig.\ref{figRheo}(b)]. In both cases, when given enough time, the sample eventually completely fluidizes: $G''>G'$ while both moduli are stationary. However, in the low stress case, the gel remains solid  until $t\sim$ 700 s whereas, in the high stress case, the gel is apparently fluidized from the start. To unveil the fluidization processes involved in those two different LAOStress experiments, we now turn to ultrasonic imaging.

%Graph2--------------------------------------------------------------------------
\begin{figure*}
\centering
\includegraphics[width=15cm]{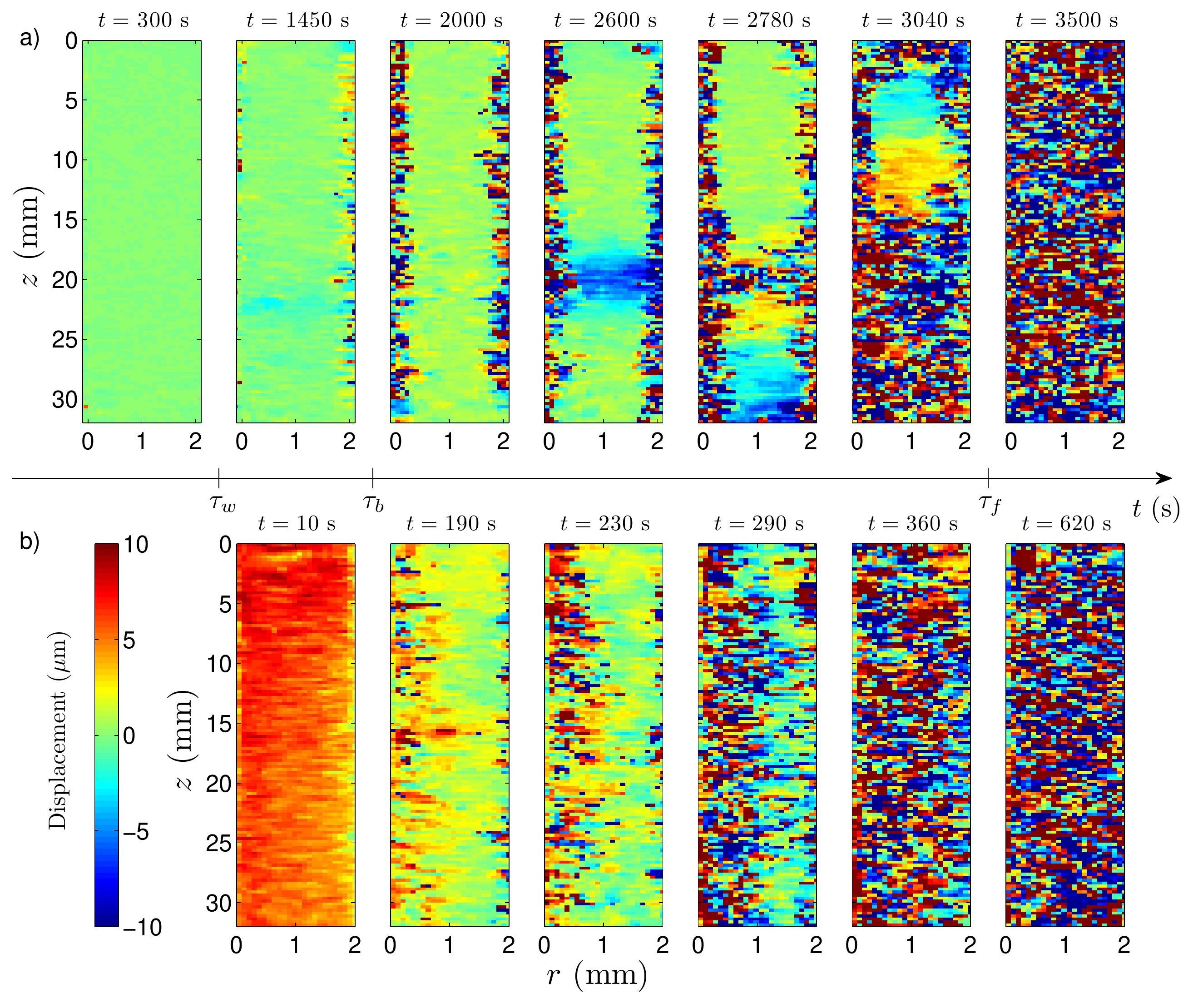}
\caption{Time evolution of the displacement field $\Delta(r, z, t)$ as measured through ultrasonic imaging during the LAOStress experiment shown in Fig.~\ref{figRheo}. The USV sampling frequency is $0.2$ Hz so that $\Delta(r, z, t)$ is probed every five oscillation cycles. The rotor is located at $r=0$ mm and the stator position is $r=2$ mm. $\Delta(r, z, t)$ is coded in linear color levels. (a)~For a {\it low} stress amplitude, $\sigma_0=15$~Pa, the gel remains completely solid ($\Delta \sim 0$) for $t<\tau_w\simeq 700$~s. For $\tau_w<t<\tau_b$ the gel is fluidized at the rotor and the stator but remains solid in the bulk. For $\tau_b<t<\tau_f$ the bulk material fluidizes heterogeneously along the vertical direction $z$. For $t>\tau_f$ the gel is homogeneously fluidized. (b)~For a {\it high} stress amplitude, $\sigma_0=27$~Pa, the gel first displays a homogeneous yet nonzero displacement field indicative of slippage at both walls. The gel subsequently shows a heterogeneous fluidization along the radial direction $r$ for $\tau_b<t<\tau_f$ and full fluidization for $t> \tau_f$. See ESI$^{\dag}$ for movies that simultaneously display the data recorded by the rheometer and the displacement field $\Delta(r, z, t)$ for both low and high stress amplitudes.}
\label{figUSV}
\end{figure*}

\subsection{Displacement field from ultrasonic imaging}
Figure \ref{figUSV} and movies in the ESI$^{\dag}$ show the evolution of the tangential displacement field $\Delta(r, z, t)$ recorded inside the Taylor--Couette device simultaneously to the rheology experiments reported in Fig.~\ref{figRheo} with $c=6\%$. One can define four successive regimes of material behaviour as characterized by different structures of the displacement field. First, for $0<t<\tau_w$, $\Delta$ is constant and equal to zero across the entire gap. This means that from one cycle to the other (or at least every five cycles since the displacement is only sampled at 0.2~Hz), the hollow glass spheres, and thus the gel material, returns into exactly the same position. Together with the fact that $G'>G''$, we interpret this as the behaviour of a homogeneous solid that adheres to the walls of the Taylor--Couette cell. This regime is only observed at low stress amplitudes, typically for $\sigma_0\lesssim 20$~Pa [see Figs.~\ref{figRheo}(a) and \ref{figUSV}(a) for $\sigma_0=15$~Pa]. For larger $\sigma_0$ one directly observes the following second regime, i.e. one effectively has $\tau_w=0$. For $\tau_w<t<\tau_b$, $\Delta$ is either zero in the bulk yet with significant values at the walls [see $t=1450$~s in Fig.~\ref{figUSV}(a)] or shows a homogeneous non-zero value in the bulk [see $t=10$~s in Fig.~\ref{figUSV}(b)]. This indicates that the gel has yielded near the walls at time $t=\tau_w$, thus creating fluidized lubrication layers at the walls although the bulk material remains solid. Consequently, we observe a plug-like flow with $\Delta \sim 0$ for low stress but non-zero, constant displacement at higher stress amplitudes, which points to a global drift of the solid-like gel from one cycle to the other and to slippage at the walls. Third, for $\tau_b<t<\tau_f$, $\Delta$ is constant over regions spanning a large amount of the cell gap yet separated by regions where $\Delta$ shows large, fluctuating and apparently random values. As already discussed in Refs.~\cite{Gibaud,Perge}, this is typical of a solid--fluid  coexistence. Indeed in fluidized zones, the hollow glass spheres seeding the dispersion show irreversible displacements from one oscillation cycle to the other due to their density mismatch with the CB gel. These displacements lead to decorrelation of successive ultrasonic speckle signals and thus to large erratic variations of $\Delta$. At low $\sigma_0$, this solid--fluid coexistence seems to occur preferentially along the vorticity direction [see $t=2780$~s in Fig.~\ref{figUSV}(a)], whereas at higher $\sigma_0$, we observe a more classical coexistence in the radial direction [see $t=230$ and 290~s in Fig.~\ref{figUSV}(b)]. The fluidized zones progressively grow and invade the whole sample until full fluidization is reached at $t=\tau_f$. Finally, the CB suspension flows like a liquid thereafter for $t>\tau_f$.

In Fig.~\ref{figTAU}, we report the three characteristic times, $\tau_w$, $\tau_b$ and $\tau_f$, for different values of the imposed stress amplitude of the LAOStress experiment determined using the ultrasonic imaging technique. Yielding was only observed in the range~$\sigma_0 \gtrsim 9$~Pa. For lower values of the stress amplitude, the sample remained solid for at least 10$^5$~s. In Fig.~\ref{figTAU}, a vertical, linear path represents the time evolution of the gel during a constant stress amplitude LAOStress experiment. Then, we can identify a region $0 < t < \tau_w(\sigma_0)$, denoted by (a) in Fig.~\ref{figTAU},  for which the sample is completely solid and adheres to the walls of the Taylor--Couette cell. In the region $\tau_w(\sigma_0) < t < \tau_b(\sigma_0)$, denoted by~(b), the sample remains solid in the bulk but slips along the walls of the cell. We associate this region (b) with a fatigue process that will lead to fluidization at a later time. In the region $\tau_b(\sigma_0) < t < \tau_f(\sigma_0)$, denoted by~(c), the material is heterogeneous exhibiting a coexistence of fluid-like and solid-like regions. The solid-like zones get progressively eroded by the fluid-like zones until the entire gel is fluidized at $t = \tau_f$. A similar erosion process has been observed in laponite gels in shear-rate controlled experiments \cite{gibaud_laponite1, gibaud_laponite2}. At both {\it low} and {\it high stresses}, $\tau_w$, $\tau_b$ and $\tau_f$ seem to follow an exponential law of type $\tau \sim e^{-\sigma_0/\sigma^*})$. The situation is complex as one can define four different $\sigma^*$ related to $\tau_b$ and $\tau_f$ at low and high stresses. In Ref.~\cite{Gibaud}, $\sigma^*$ at high stresses associated with $\tau_f$ is interpreted as the stress barrier necessary to be overcome so that the gel yields. In the creep experiments of Ref.~\cite{Sprakel}, a similar behaviour is observed and $\sigma^*$ associated with $\tau_f$ is modeled in the delayed yielding framework \cite{Lindstrom}. In this model, the change of slope between the low and high stress regime is attributed to a change of the average number of particles involved in the weakest link of the gel network mesh. We believe that applying the delayed yielding model on $\tau_f$ is however inappropriate and that $\tau_f$ corresponds here to an accumulation of multiple yielding processes. 

\subsection{Normal forces}%--------------------------------------------------------------------------
\begin{figure}
\centering
\includegraphics[width=7.5cm]{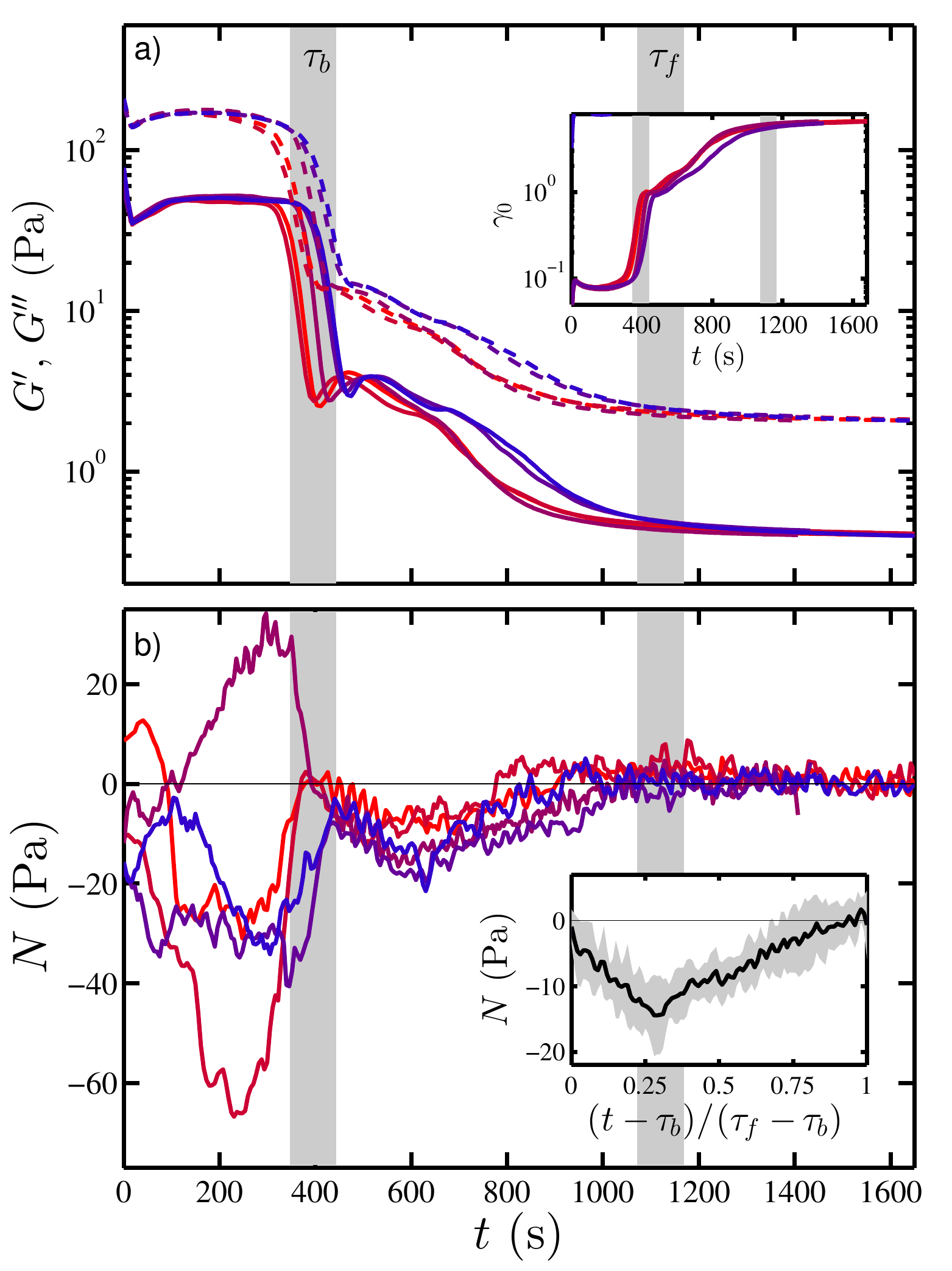}
\caption{Time evolution of the normal force during a LAOStress experiment for a carbon black gel at $c=$6\%w/w and $f=1$~Hz in a cone--plate geometry at $\sigma_0=20$~Pa. We have repeated the experiment 5 times to check for reproducibility. (a)~Time evolution of the elastic modulus ($G'$, solid line) and loss modulus ($G''$,  dashed line). The insets show the time evolution of the strain amplitude $\gamma_0$. The behaviour of $G'$, $G''$ and $\gamma_0$ is similar to the one observed for the Taylor--Couette geometry (Fig.~\ref{figRheo}(b)). (b)~Time evolution of the normal force $N$. Inset: average value of $N$ over the 5 experiments for $\tau_b<t<\tau_f$. The shaded area corresponds to the standard deviation.}
\label{figN}
\end{figure}

On top of this complex spatiotemporal yielding behaviour, we observe the presence of normal forces. In Fig. \ref{figN}, we measure the normal force $N$ in the {\it high stress} regime during a LAOStress experiment in a cone--plate geometry. We observe that $G''>G'$ at $t=0$ s: the gel directly yields at the wall, which is consistent with the high stress experiments in the Taylor--Couette geometry.  We observe dramatic irreproducible fluctuations of the normal force. In our opinion, the value of the normal force might be related to residual stress within the gel \cite{Osuji,Grenard}. However, as soon as the gel partially fluidizes in the bulk, for $\tau_b<t<\tau_f$, we consistently measure a negative normal force. This effect is robustly observed in all experiments as we varied the frequency $f$ or the imposed stress amplitude $\sigma_0$.

To understand the effect of the normal forces on the erosion process during the solid--fluid coexistence, we assume that deformations within the gel are small, and we use the linearized theory of elasticity for the strand network which is assumed to be isotropic. The gel is supposed uniform but partially fills the gap. Such a model is derived is Appendix 1 and demonstrates how normal forces directly affect the stress state within the solid fraction of the solid--fluid coexistence. We find that the presence of a negative normal force increases the maximum principal stress within the gel. Therefore, if the normal force is independent of the applied stress $\sigma_0$, the correction to $\sigma_0$ induced by the presence of normal forces only shifts $\tau_f$ to lower stresses but does not affect the scaling.

%Graph4--------------------------------------------------------------------------

\subsection{Scaling of the characteristic times}%--------------------------------------------------------------------------
\begin{figure*}
\centering
\includegraphics[width=12.5cm]{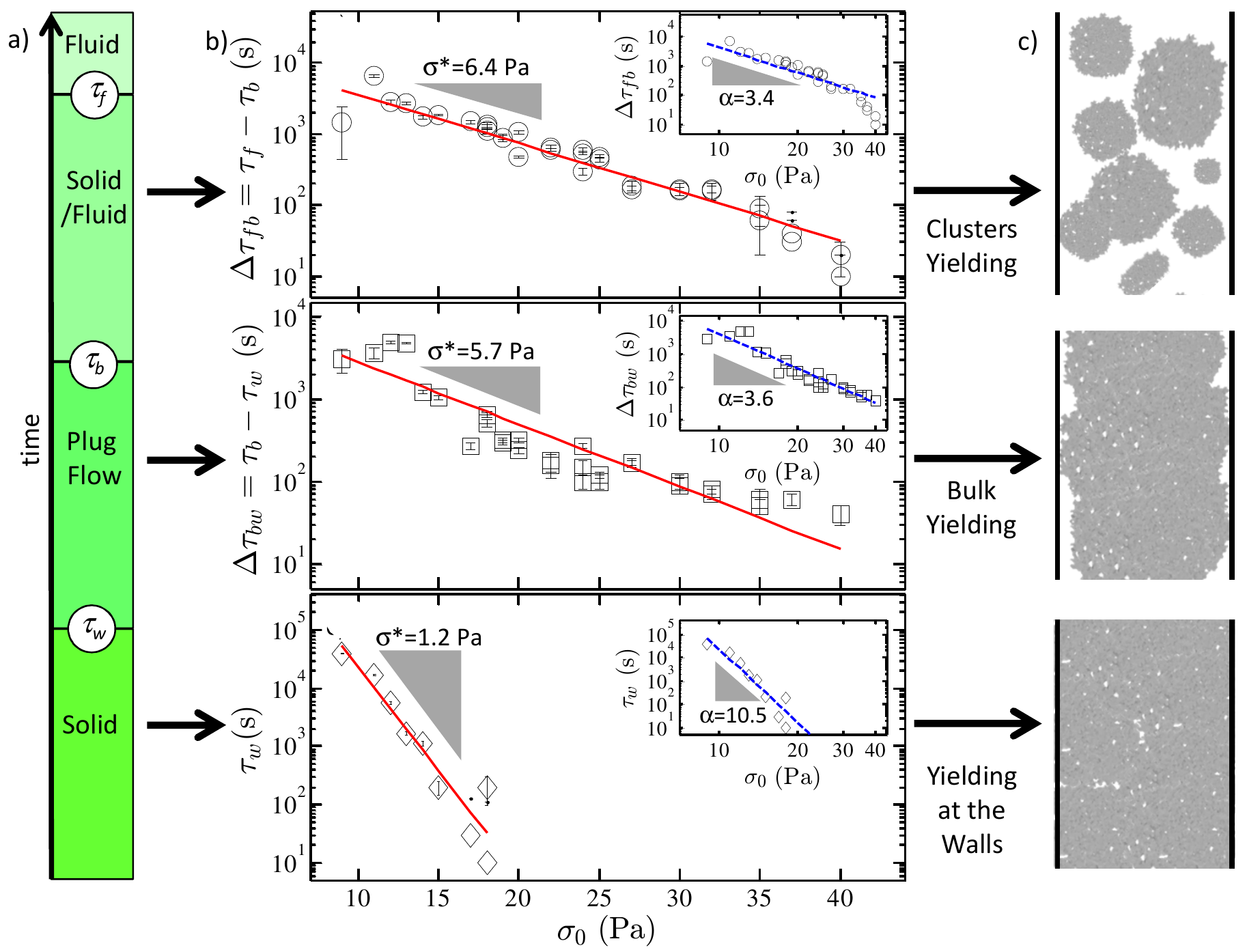}
\caption{Scaling properties of the characteristic durations as a function of the applied stress amplitude $\sigma_0$. (a)~Flow state as a function of time. (b)~$\tau_{w}$, $\Delta\tau_{bw}$ and $\Delta\tau_{fb}$ vs $\sigma_0$. Red solid lines are exponential fits ($\Delta\tau=Ae^{-\sigma_0/\sigma^*}$) to the data whereas blue dashed lines in the insets are power-law fits ($\Delta\tau=B\sigma_0^{-\alpha}$) to the same data. (c)~Sketch of the gel structure between the stator and the rotor.}
\label{figTAUs}
\end{figure*}

Having excluded the normal forces to explain the different behaviour at low and high stresses, we turn our attention to the duration of each yielding process. We note that at {\it low stresses}, all the characteristic times have a similar value $\tau_w \sim \tau_b \sim \tau_f$. However, at {\it high stresses} the presence of lubrication layer seems to split those characteristic times apart from one another. Given the fact that in addition to $\tau_f$, we have $\tau_b$ and $\tau_w$, we can define the duration of each event and compare the {\it low stress} and {\it high stress} regimes. $\tau_w$ corresponds to the duration over which the gel remains entirely solid. $\Delta\tau_{bw} = \tau_b - \tau_w$ is the duration of the fatigue regime when the gel slips at the wall but remains solid in bulk. $\Delta\tau_{fb} = \tau_f - \tau_b$ is the duration of solid--fluid coexistence. As shown in Fig. \ref{figTAUs}, the discrepancy between low and high stresses vanishes when those three durations are plotted as functions of the stress amplitude $\sigma_0$. We have therefore fitted the data with a single exponential or a single power law for each one of the three durations. Given the narrow range of experimentally accessible values of $\sigma_0$, it is difficult to discriminate between exponential and power-law fits (see Table~\ref{tab_1} for the values of the best fit parameters in both cases). Still, this scaling allows us to distinguish three successive yielding phenomena: yielding at the wall, bulk yielding and clusters yielding. The gel hierarchically yields in time and each successive yielding process requires a higher stress barrier $\sigma^*$ or a lower exponent $\alpha$.
%Graph3--------------------------------------------------------------------------
\begin{figure}
\centering
\includegraphics[width=7.5cm]{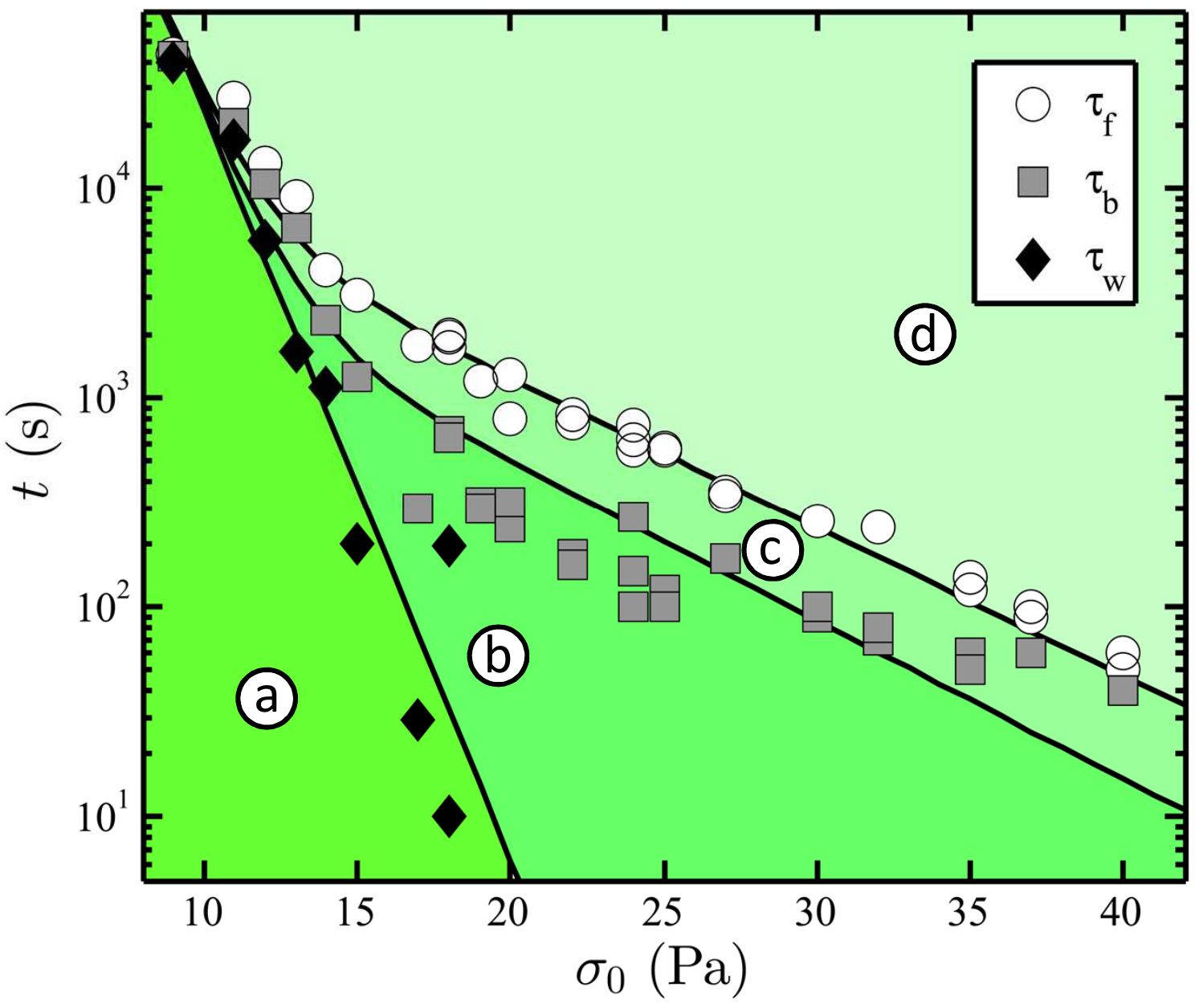}
\caption{Dynamic state diagram of the fatigue process of a carbon black gel at $c=6\%$~w/w under LAOStress at a frequency of $f=1$~Hz. The characteristic times $t = \tau_{w}$ ($\blacklozenge$), $t = \tau_{b}$ ($ \blacksquare$) and $t = \tau_{f}$ ($\circ$) are determined from ultrasonic imaging and define the boundaries between different states encountered during the fatigue process: (a) solid-like state, (b) plug-like flow i.e. solid-body displacement with slip at the walls, (c) solid--fluid coexistence along the vorticity direction, and (d) a fluid-like state. Solid lines are exponential fits based on Table~\ref{tab_1}.}
\label{figTAU}
\end{figure}

\section{DISCUSSION}%--------------------------------------------------------------------------
\label{s.discuss}

\subsection{Basquin model and delayed failure model}%-------------------------------------------------------------------------
Gaining insight into the physical principles governing our experiments from a theoretical perspective remains a challenge. The entire process shows multiple yielding time scales, which suggests that many mechanisms are at play. The gel first yields at the wall: interaction between the wall and the gel should thus be taken into account. The gel then yields heterogeneously in the bulk, producing large clusters that eventually fluidize, which suggests that a mean-field theory is inappropriate. Keeping those difficulties in mind, and given the fact that no theoretical approach currently responds to all those criteria, we examine two models: the Basquin model and the delayed failure model. Both models rely on a mean-field approach. We shall therefore apply them on homogeneous processes only, such the duration $\tau_{w}$ of the initial solid regime and that of the fatigue regime $\Delta \tau_{bw}$.

%Graph5--------------------------------------------------------------------------
\clearpage

The power-law behaviour of the yielding characteristic times is reminiscent of the empiric Basquin law~\cite{Basquin}, where $\tau_f \sim \sigma_0^{-\alpha}$. This law can be derived from a fiber bundle model \cite{Kun}. The gel is then represented by a set of parallel fibers organized on a regular lattice. When the bundle is subjected to an increasing external load, the fibers behave like linear elastic springs until they break for a given failure load. To recover the Basquin law, Kun et al.~\cite{Kun,Halasz} modified the fiber bundle model and considered a mean-field approach in which the network strands form a random network and fail either due to immediate breaking or to aging through damage accumulation. To capture damage recovery due to healing of micro-cracks or thermally activated rebinding of failed contacts, Kun et al.\cite{Kun,Halasz} introduced a memory term which scales exponentially with a characteristic time of the system. In this framework, $\alpha$ is directly related to the growth law of local damage as a function of the local stress. High values of $\alpha$ mean that the material is prompt to accumulate damage. Fits of the experiments give $\alpha_w=10.5$ and $\alpha_{bw}=3.6$ for $\tau_{w}$ and $\Delta \tau_{bw}$ respectively (see Table \ref{tab_1}). The value of $\alpha_{bw}$ is close to the Basquin exponents found in recent creep experiments on carbon black gels ($\alpha\sim 2$--3) \cite{Grenard}, carbopol microgels ($\alpha\sim 3$--8) \cite{Divoux2} and casein biogels ($\alpha\sim 5$) \cite{Leocmach} but very high when compared to metals ($\alpha\sim 0.1$) \cite{Eshbach} and to asphalt ($\alpha\sim 0.5$) \cite{Kun,Halasz}. Soft gels are indeed prone to accumulate damage and therefore they are much more sensitive to stress than metals. We also note that our data show no hint of a critical stress as measured in Refs.~\cite{Grenard,Divoux2}.

%table1--------------------------------------------------------------------------
\begin{table}
\centering
\begin{tabular}{c c c c c } \hline
& $A$ (10$^4$ s) & $\sigma^*$ (Pa) & $B$ (10$^7$ Pa$^{\alpha}$s) & $\alpha$ \\ \hline
$\tau_{w}$ & 9$\pm$1 10$^3$ & 1.2$\pm$0.2 & 8$\pm$1 10$^7$ & 10.5$\pm$1.1 \\ \hline
$\Delta\tau_{bw}$ & 1.6$\pm$0.2& 5.7$\pm$1.1 & 1.0$\pm0.1$ & 3.6$\pm$0.5 \\ \hline
$\Delta\tau_{fb}$ & 1.7$\pm$0.2 & 6.4$\pm$1.0 & 3.1$\pm0.3$ & 3.4$\pm$0.4 \\ \hline
\end{tabular}
\caption{Parameters used in Fig. \ref{figTAUs} to fit the data by an exponential ($\Delta\tau=Ae^{-\sigma_0/\sigma^*}$) and by a power law ($\Delta\tau=B\sigma_0^{-\alpha}$) }
\label{tab_1}
\end{table}

To model the exponential behaviour of the characteristic times of the yielding gel, we turn to the delayed failure model~\cite{Lindstrom}. This is also a mean field approach, based at the micro-scale on Kramers' transition state theory~\cite{Kramers}. Here, the particles form percolating strands where $n$ is the average number of colloids in the  cross-section of the strand at its weakest point. When no stress is applied, thermal fluctuations are assumed to dissociate or associate interparticle bonds at rates $k_D$ and $k_A$ respectively, with $k_D < k_A$ for attractive gels. A constant $C$ with units of compliance is also postulated, which converts the applied stress $\sigma$ into work $C\sigma$ with units of $k_B T$. This work lowers the interaction potential across each bond. Two strand-breaking processes were then proposed: one for high-stress case, leading to sequential bond-breaking within a strand and an activation stress $\sigma^* = 1/C$, and one for the low-stress case which allows for re-association of bonds (healing) within the strand and an activation stress $\sigma^* = 1/nC$. Both bond-breaking processes lead to an exponential scaling of the time-to-failure. In Appendix 2, we adapt the theoretical frame developed for failure of colloidal gels under a constant load to cyclic loading. As for creep experiments, the delayed failure model for cyclic loading yields an exponential behaviour of the characteristic failure time $\tau^{mod}$ with respect to $\sigma_0$: $\Delta\tau=Ae^{-\sigma_0/\sigma^*}$. However, the prefactor $A$ to this exponential is greater than that of the creep, and leads to a longer failure time: gels resist better to oscillations than to creep for the same activation stress $\sigma^*$ and applied load $\sigma_0$. For $\Delta\tau_{bw}$, we find $\sigma^*\approx 5.7$. In this model $n$ and $C$ are coupled and cannot be measured independently; we take values similar to those of Ref.~\cite{Lindstrom}, $k_D=0.26$~s$^{-1}$, $k_A=1.5$~s$^{-1}$, $1/C=28.5$~Pa and $n=5$ as benchmarks for further discussion. In the case of $\tau_{w}$,  $\sigma^*\approx 1.2$ is much smaller, implying that the gel is weaker in the wall region. Keeping $C$ constant, it requires to divide $n$ by $\sim$5 which means that the CB particle strands develop fewer links with the walls as compared to the bulk. Conversely, keeping $n$ constant, it requires to divide $1/C$ by about 5 which means that the CB particles have a much weaker interaction with the walls than between themselves. 

Which model is best suited to interpret the present experiments remains an open question. The fiber bundle model can solely support longitudinal deformation which allows for studying only loading of the bundle parallel to fibers so that its use to model our experiment performed under shear is at best qualitative. Moreover, the Basquin law stipulates that the gel will always fail even at very low stresses. Experimentally, it seems that below roughly 9~Pa the gels remains solid, even though it cannot be excluded that it fails after some very long time exceeding experimental time scales. In the delayed yielding model, a divergence of the time-to-failure can be explained through the dominance of healing in the competition between healing and rupture of strands. A predictive capability for this critical stress is not yet available. We also note that $\sigma^*=6\pm$1~Pa roughly matches the stress below which it seems impossible for fatigue to eventually fluidize the gel. 

\subsection{Influence of the frequency of the LAOStress}%--------------------------------------------------------------------------
\begin{figure}
\centering
\includegraphics[width=7.5cm]{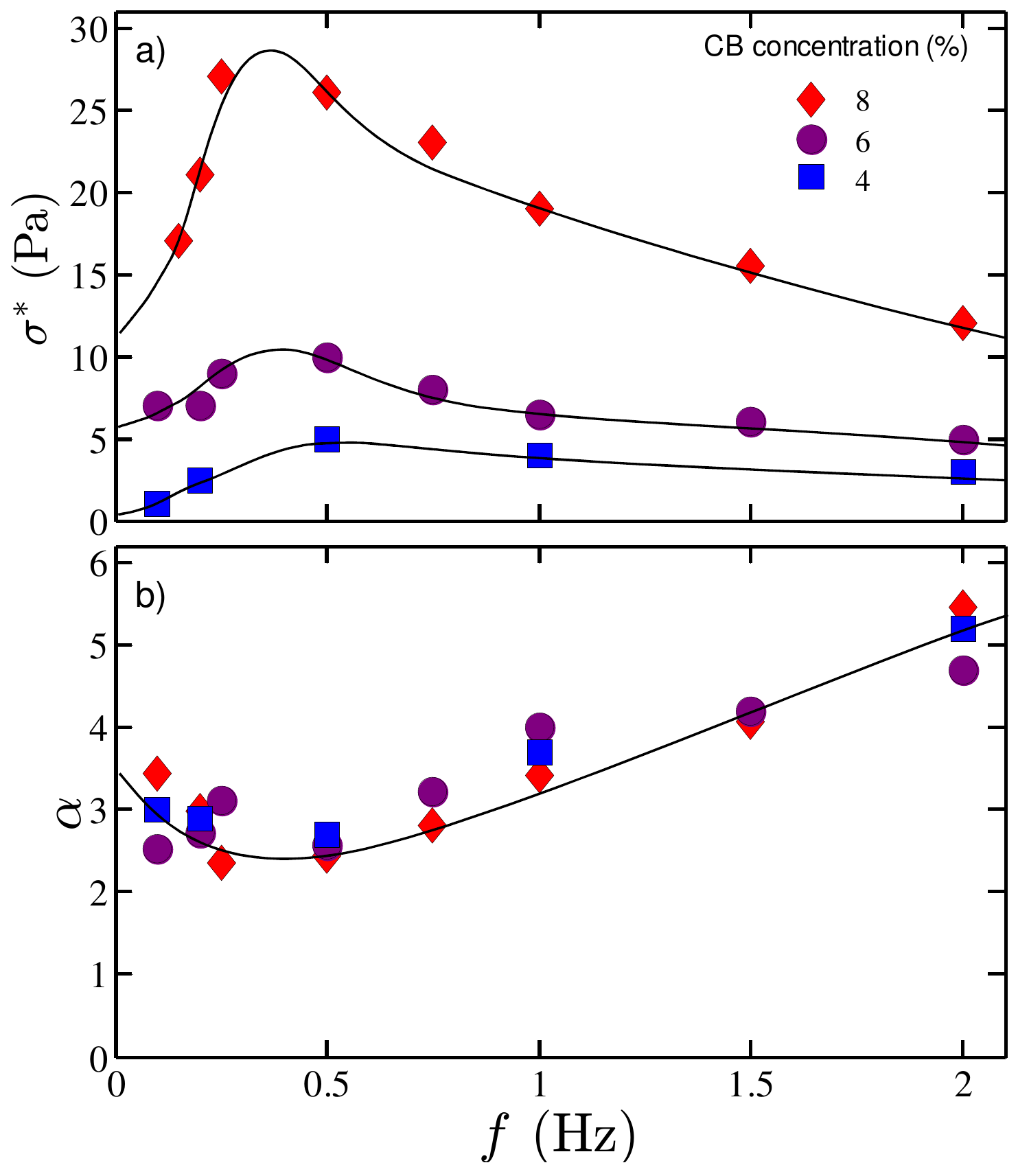}
\caption{Evolution of (a) the stress barrier $\sigma^*$ and (b) the power-law exponent $\alpha$ with the oscillatory stress frequency $f$ for a carbon black gel at $c=4$ ($\blacksquare$), 6 ($\bullet$) and 8\% w/w ($\blacklozenge$). Lines are included to guide the eye.}
\label{figSIG}
\end{figure}

So far we have discussed the measurements and the interpretation of the scaling behaviour of the characteristic times with $\sigma_0$. We now focus on two parameters that are likely to modify the scaling parameters of $\tau_f$ with $\sigma_0$, namely the CB concentration $c$ and the oscillation frequency $f$. In general, $f$ and $c$ do not qualitatively affect the fatigue scenario described in Fig. \ref{figTAU}. Moreover, in the high stress regime, the $\tau_f(\sigma_0)$ data are well fitted both by an exponential and by a power law. Figure~\ref{figSIG} shows the values of $\sigma^*$ and $\alpha$ as a function of $f$ and $c$. 

As for exponential fits, the stress barrier $\sigma^*$ increases sharply with $c$. This seems reasonable since a concentration increase reinforces the backbone structure of the gel. On the contrary, the exponent $\alpha$ found from power-law fits does not display any systematic evolution with $c$. Although we probe higher $\sigma_0$ for higher $c$, the susceptibility of the gel appears to be independent of $c$ within the fiber-bundle model interpretation of the Basquin law.

The major advantage of oscillatory experiments compared to creep experiments \cite{Gibaud,Grenard} is that LAOStress is frequency-resolved. Figure~\ref{figSIG}(a) shows that $\sigma^*$ first increases with $f$ at low frequencies, then reaches a maximum around $f$=0.4$\pm$0.15 Hz and finally decreases for higher frequencies. The delayed yielding model does not predict any frequency dependence of $\sigma^*$ (see Fig. \ref{nico}(a) in Appendix~2). Yet, we know that the gel rebuilds it-self on time scales that are below 1~s \cite{Gibaud}. Therefore decreasing the frequency should leave more time for the gel to heal between two successive oscillation cycles and make the gel more resistant to stress. This could explain the decreasing curve $\sigma^*(f)$ at high frequencies. On the other hand, $\alpha$ displays a minimum around $f$=0.4$\pm$0.15 Hz while the Basquin model also does not predict any frequency dependence since it only depends on the number of cycles. Yet the increase of $\alpha$ at high frequencies is qualitatively consistent with the previous healing argument that leads to increased susceptibility to stress as $f$ increases. We note that Z. Shao et al. \cite{Shao} have also observed a puzzling, nonmonotonic frequency dependence in the yielding process of flocculated microgel suspensions. Currently, we do not have any sensible argument to interpret the nonmonotonic variations of $\alpha$ or $\sigma^*$ with $f$ as reported in Fig.~\ref{figSIG}, and more specifically their low-frequency behaviour.

%Graph6--------------------------------------------------------------------------

\section{CONCLUSION}%--------------------------------------------------------------------------
We have investigated the fatigue scenario in LAOStress experiments on carbon black gels. Yielding was only observed within experimental time scales when the applied stress amplitude exceeded a critical value ($\sigma_0 \gtrsim 9$~Pa for $c = 6\%$~w/w). Above this applied stress, we have determined three characteristic times for delayed yielding under LAOStress based on ultrasonic imaging: within the time interval $0<t<\tau_w$ the gel remains solid, within $\tau_w<t<\tau_b$ the gel yields at the walls but remains solid in the bulk, and within $\tau_b<t<\tau_f$, we observe solid--fluid coexistence associated with negative normal forces. Beyond $\tau_f$, the sample flows like a liquid. Those characteristic times scale differently with the applied stress $\sigma_0$ at low and high stresses. However this difference vanishes when we scale the characteristic durations of each process $\tau_w$, $\tau_b-\tau_w$, $\tau_f-\tau_b$ with $\sigma_0$. Such a scaling allows us to distinguish three successive yielding processes: yielding at the wall, bulk yielding and clusters yielding.

Those durations are functions of the applied stress and can be empirically fitted by an exponential, although a power law cannot be ruled out due to the limited range of the stress amplitude.
To gain insight into those multiple yielding processes, we have used the delayed yielding model extended to LAOStress ($\tau \sim \exp(-\sigma_0/\sigma^*)$)  and the Basquin model ($\tau \sim \sigma_0^{- \alpha}$). Both models fit reasonably well the data. This analysis shows that each successive yielding process requires a higher stress barrier, $\sigma^*$ or a lower exponent $\alpha$. According to the Basquin model, our gels are prompt to accumulate damage and are much more sensitive to stress than asphalt \cite{Kun} or metals \cite{Eshbach}. According to the delayed failure model, the carbon black particles have a weaker interaction with the walls than between themselves which justifies that the gel yields first at the wall and then in bulk.

The yielding process also depends on the concentration of the gel and on the applied frequency in a complex manner that remains to be fully understood. Another interesting line of enquiry, left for future study, is the effect of the particle--wall interactions on the yielding at the wall. Presumably, weak or repulsive particle wall interactions would reduce the time-to-failure $\tau_w$ at the wall, while strong attractive particle--wall interactions could prolong $\tau_w$, or even re-localize initial yielding to the bulk. In the same spirit the influence of surface roughness, which has already been addressed in carbon black dispersions submitted to a steady shear stress\cite{Grenard}, should also be investigated and modelled in the case of LAOStress.

Finally, there is some similarity between the cyclic fatigue of the colloidal gel and the very high cycle fatigue (VHCF)~\cite{Pyttel} observed for metal alloys. When subjected to a cyclic stress far below the yield stress, microcracks develop dispersed within the crystal grains. These microcracks multiply and grow until they reach the grain boundaries. This triggers an avalanching crack growth and ultimate macroscopic failure. Similarly to carbon black gels, the time-to-failure for VHCF in metals decays exponentially with the stress amplitude. Hence, despite the very different microstructures of metals and colloidal gels, there are striking similarities in their fatigue behaviour that presumably derive from thermally-activated microscale processes governing irreversible damage.

\section*{Appendix 1: Normal forces model  }%--------------------------------------------------------------------------
To potentially understand the origin of the normal forces during the solid--fluid coexistence, we assume that deformations within the gel are small and we use the linearized theory of elasticity for its strand network. The strand network is further assumed to be isotropic. We take the gel to be in a state of uniform, plane strain, with zero deformation in the $z$~direction of the Taylor--Couette geometry or the normal direction of the cone--plane geometry. We also neglect the curvature of these geometries. Thus, we introduce a rectangular coordinate frame, with base vectors $\{\vec{e}_1, \vec{e}_2, \vec{e}_3\}$, so that $\vec{e}_1$ is the shear direction and $\vec{e}_2$ is the shear gradient direction. From the plain strain geometry, we have~$\sigma_{13}=\sigma_{31}=\sigma_{23}=\sigma_{32} = 0$. The remaining stress components depend on the boundary conditions.
When the gel region fills the gap of the geometry and adheres to the walls, the boundary condition is
\[ \sigma_{12} = \sigma(t), \]
while we have $\epsilon_{11} = \epsilon_{22} = \epsilon_{33} = 0$ inside the gel, where $\bar{\bar{\epsilon}}$ denotes the Euler--Almansi strain tensor. The above yields the well-known simple shear solution
\[ \sigma_{12} = \sigma_{21} = \sigma(t), \]
while all other stress components vanish for infinitesimal strains. The principal tensile stress then becomes identical to the applied shear stress~$\sigma_{\mathrm{p}}(t) = \sigma(t)$.

During solid--fluid coexistence, the gel only fills a fraction of the gap, and the boundary conditions on the gel become
\[ \sigma_{12} = \sigma(t),\quad \sigma_{22} = -N(t), \]
where $N(t) $ is the normal force observed in experiments.  From the geometry, $\epsilon_{11} = \epsilon_{33} = 0$ inside the gel. Hooke�s law for plane strain and the equilibrium equation for an isotropic, linear elastic material in plain strain, then gives
\begin{subequations}
\begin{eqnarray}
(1-\nu) \sigma_{11} - \nu \sigma_{22} & = & 0 
\\
\sigma_{33} - \nu (\sigma_{11} + \sigma_{22}) & = & 0 
\end{eqnarray}
\end{subequations}
where $\nu$ is the Poisson modulus of the gel. The stress tensor becomes
\begin{equation}
\bar{\bar{\sigma}} = \left[ \begin{array}{c c c} - \frac{\nu}{1-\nu} N(t) & \sigma(t) & 0 \\
\sigma(t) & - N(t) & 0 \\ 0 & 0 & - \frac{\nu}{1-\nu} N(t) \end{array} \right].
\end{equation}
The principal tensile stress is obtained by computing the maximum eigenvalue of~$\bar{\bar{\sigma}}$, which gives
\[ \sigma_{\mathrm{p}}(t) = - \frac{1}{2(1-\nu)} N(t) + \sqrt{ \frac{1-4\nu(1-\nu)}{4(1-\nu)^2} N^2(t) + \sigma^2(t)}. \]
Therefore, the presence of a negative normal force increases the principal stress on the gel and vice versa.

\section*{Appendix 2: Delayed failure model for cyclic loading }%--------------------------------------------------------------------------
The delayed failure model is a mean field approach that describes the dynamics of the fraction of remaining strands, $x$, in a yield surface of a colloidal gel subject to a creep experiment at constant stress $\sigma(t) = \sigma_0$. At start, the gel is completely solid and $x=$1. The dynamics is then driven by type ordinary differential equation (ODE) of birth/death type~\cite{Lindstrom}:
\begin{equation}\label{eqn:dissrate}
\frac{d x}{d t} = - K_D(x,\sigma(t) ) x + K_{A}(1-x);
\end{equation}

$K_D(x,\sigma_0)=K_0\exp(\frac{nC\sigma_0}{x})$ is the dissociation rate of the strands and $K_A$ is their association rate. $K_D$ depends on local parameters such as $k_A$ and $k_D$, respectively, the association and dissociation rate of the individual colloids, $n$ the average number of colloids in the cross section of a strand at its weakest point, the temperature $T$, the range of the attraction between colloids $\delta$, and the initial area density of strands $\rho_0$. Indeed, the exponential behaviour of $K_D$ is related to an activated process where the stress barrier necessary to break a strand composed of $n$ colloids in its cross section is $1/nC=\rho_0k_BT/\delta$. The complete expression of the amplitude of $K_D$ can be found in Ref.~\cite{Lindstrom}.

Instead of looking at a creep experiment where $\sigma(t)=\sigma_0$ we now consider LAOStress experiments where the local stress on the strands is taken to be $\sigma(t) = \sigma_0|\sin 2\pi f t|$ in Eq.~\eqref{eqn:dissrate}. Taking $\sin 2\pi f t$ instead of $|\sin 2\pi f t|$ would lead to $K_D \sim 0$ when the $\sin$ function becomes negative, which does not correspond to the actual situation where the sample experiences the same dissociation rate independently of the sign of the applied shear stress. We then discretize $x$ into $x_m=x(m/f)$, the fraction of the strands in the gel at the $m^{th}$ cycle, $m$ being an integer. 
Integrating Eq.~\eqref{eqn:dissrate} over the $m^{\mathrm{th}}$ loading cycle with respect to time, while assuming that $x$ changes only marginally during each cycle, gives a recurrence relation

\begin{eqnarray}
x_{m+1}-x_m&=& - x_m \int_{m/f}^{(m+1)/f} \!\!\!\! K_0\exp\left( \frac{nC \sigma_0}{x_m} |\sin{2\pi f t}|\right) \mathrm{d} t \nonumber\\& &+ \frac{K_{A}}{f} (1-x_m) \nonumber
\\
&=& -\frac{ K_0}{f} x_m \Upsilon \left( \frac{nC \sigma_0}{x_m} \right) + \frac{K_{A}}{f} (1-x_m), \label{eqn:recurr}
\end{eqnarray}
with initial value~$x_0=1$, and where
\begin{equation}
\Upsilon(a) = 2 \int_0^{1/2} \exp(a|\sin{2\pi s}|) d s.
\end{equation}
Using again that $x_{m+1} - x_m$ is small, the recurrence equation~\eqref{eqn:recurr} can be approximated by a differential equation, now viewing~$m=ft$ as a real number:
\begin{equation}\label{eqn:fatigueode}
\frac{d x}{d t} = f \frac{d x}{d m} = - K_0\Upsilon \left( \frac{nC\sigma_0}{x} \right) x + K_{A}(1-x).
\end{equation}

This ODE describes how distributed damage evolves under cyclic loading conditions. To the knowledge of the authors, no numerical solution exists to Eq.~\eqref{eqn:fatigueode}.
For a sufficiently large $K_{A} > 0$, that is if ruptured strands are assumed to re-associate, the numerical solutions show that $x$ converges toward a non-zero equilibrium value related to the ratio $ K_{A}/ K_0$:  no failure can be predicted with certainty. Still, experimentally we always observe a binary behaviour of the yielding process: either $\sigma_0$ is too small and the suspension remains solid, $x=1$ over some seriously long time, or the suspension eventually completely fluidizes, $x=0$. We therefore restrict the following discussion to the fluidization process and assume that $ K_{A}=0$.

\begin{figure}
\centering
\includegraphics[width=7.5cm]{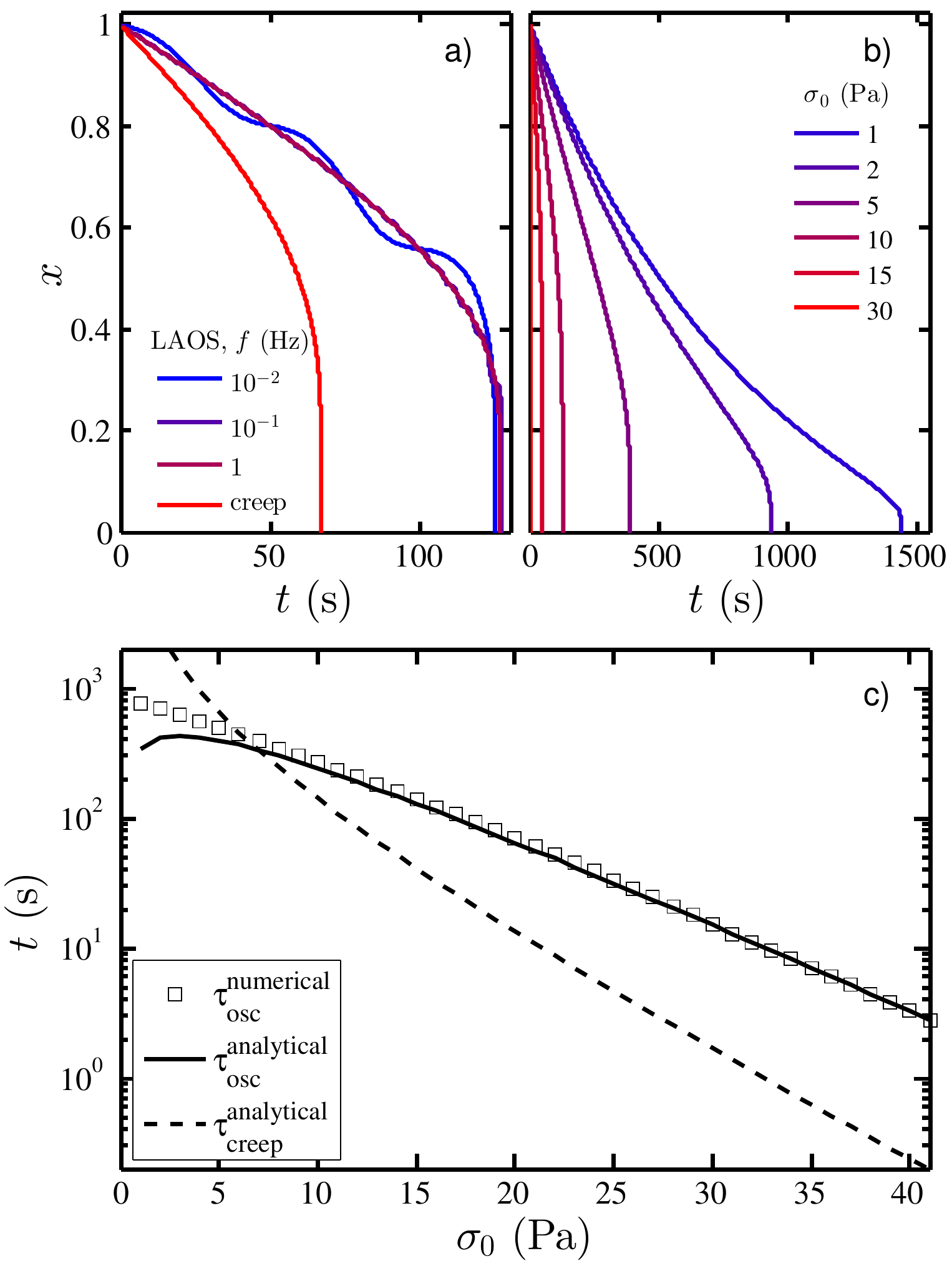}
\caption{Resolution of the delayed failure model for LAOStress. (a)~Density of strands $x$ remaining solid under LAOStress as a function of time for different frequencies at a constant applied stress amplitude $\sigma_0=6$ ~Pa. (b)~Density of strands $x$ remaining solid under LAOStress as a function of time for different applied stresses at $f=1$~Hz. (c)~Characteristic times obtained by solving Eq.~\eqref{eqn:fatigueode} numerically $\tau_{osc}^{numerical}$ ($\square$), and Eq.~\eqref{eqn:fatigueodesimplified} analytically  ($-$). $\tau_{creep}^{analytical}$ ($--$) is the analytical solution of  the delayed failure model for the corresponding creep experiments given in Eq.~\eqref{eqn:creepmodel}. The parameters of the model are displayed in Table \ref{tab_nico}.}
\label{nico}
\end{figure}

The numerical solution of the ODE in Eq.~\eqref{eqn:dissrate} with $ K_A=0$ gives the strand density remaining attached $x$ as a function of various parameters such that $K_0$, the oscillation frequency $f$ and the amplitude of the applied stress $\sigma_0$. As shown in Fig.~\ref{nico}, $x$ decreases from 1 to 0 as a function of time: the gel gets completely fluidized. In this model, provided that the oscillation inverse frequency is small compared to the yielding time, the frequency does not influence the yielding process [see Fig.~\ref{nico}(a)]. $\sigma_0$ sets the timescale of the fluidization process [see Fig.~\ref{nico}(b)]. We extract two characteristic times from this numerical resolution: the time $\tau_{fluid}=t(x=0)$ at which $x=0$ and the characteristic yielding time $\tau_{osc}$ defined as the inverse slope of the tangent to $x$ at $t$=0. $\tau_{osc}$ is the time that matches the closest the definition of the experimental duration $\tau_{w}$, the time necessary for the gel to yield at the wall or $\Delta\tau_{bw}$, the duration necessary for the gel to yield in bulk. Indeed, in both those regimes and in the region of interest $x$ is close to 1. Note that $\tau_{fluid}$ should not be compared to $\tau_f$ as the model assumes an homogeneous yielding of the strands; experimentally we know that this is not the case. The model is thus only optimal to describe the experiments close to $x=1$ when the gel is solid and Fig.~\ref{nico}(c) only shows $\tau_{osc}$ vs $\sigma_0$.

%Graph5--------------------------------------------------------------------------

%table2--------------------------------------------------------------------------
\begin{table}
\centering
\begin{tabular}{ r c c c c c } \hline
& $n$ & $C$ (Pa$^{-1}$) & 1/$nC$ (Pa) & $k_D$ (s$^{-1}$) & $k_A$ (s$^{-1}$) \\ \hline
$\Delta \tau_{bw}$ & 5 & 0.033 & 6 & 0.26 & 1.5 \\ \hline
\end{tabular}
\caption{Parameters used in the numerical resolution of the delayed failure model for LAOStress and chosen so as to roughly match the values of $\Delta\tau_{bw}$ in Fig.~\ref{figTAUs}. Note that different sets of parameters can give the same solutions to the ODE. Indeed $n$ and $C$ are coupled and as long as the stress barrier $1/nC$ is $constant$, the solutions are identical.}
\label{tab_nico}
\end{table}

Although no analytic solution is available for Eq.~\eqref{eqn:fatigueode}, we can find a solution for the limit $x$ close to 1. The  solutions of Eq.~\eqref{eqn:fatigueode} crucially depends on the properties of~$\Upsilon$. For strong gels~\cite{Lindstrom}, like carbon black gels, failure is only observed at experimental time-scales if $nC\sigma_0 \gg 1$. Therefore, the asymptotic behaviour of~$\Upsilon$ for large~ $\sigma_0$ is pivotal. For large values of $\sigma_0$, the integral defining~$\Upsilon$ is dominated by the short interval near $s=1/4$ where the sine function has its maximum. Consequently, we may use a Taylor expansion for~$\sin 2\pi s$ around~$s=1/4$ and change the limits of the integral to~$\pm\infty$, this way obtaining

\begin{equation}
\Upsilon(a) \approx 2 \int_{-\infty}^{\infty} \exp\left\{ a \left[1-\frac{(2\pi s-1/4)^2}{2}\right] \right\} d s = \sqrt{\frac{2}{\pi a}}e^a,
\end{equation}
 for $a \gtrsim 1$. Using this approximation with Eq.~\eqref{eqn:fatigueode} gives 

\begin{equation}\label{eqn:fatigueodesimplified}
\frac{d x}{d t} 
= - K_0 \sqrt{\frac{2}{\pi nC \sigma_0 }} 
e^{  \frac{nC\sigma_0}{x}} x^{3/2} + K_{A}(1-x),\quad x(0) = 1.
\end{equation}

Given that the solution is reminiscent of an exponentially decaying function, we obtain a characteristic time-scale for fatigue failure of strong gels with $K_A=0$, simply by taking
\begin{equation}
\tau_{osc}=\left( \frac{x}{\frac{d x}{d t}} \right) _{t=0}=\frac{\sqrt{2/\pi}}{K_0} (nC\sigma_0)^{1/2}e^{-nC\sigma_0}, \quad nC\sigma_0 \gg 1.
\end{equation}
This can be compared to the corresponding time in a creep experiment, $\tau_{creep}$, which was previously found to be~\cite{Sprakel,Lindstrom}
\begin{equation}\label{eqn:creepmodel}
\tau_{creep} = \frac{1}{K_0} (nC\sigma_0)^{-1} e^{-nC\sigma_0}.
\end{equation}
The numerical solutions to Eq.~\eqref{eqn:fatigueode} with $K_A=0$ compare excellently with the analytical solution of Eq.~\eqref{eqn:fatigueodesimplified} when sampled at the frequency~$f$, $nC\sigma_0\gg$1 and $x$ close to 1 [see Fig.~\ref{nico}(c)].

To conclude, the delayed yielding model shows that the characteristic failure time varies exponentially with $\sigma_0$ for both LAOStress and creep experiments: $\tau_{osc}\sim\tau_{creep} \sim\exp({-\sigma_0/\sigma^*})$. The intimate relation between creep and cyclic fatigue originates from the common microscopic process governing failure at low stresses in the case of this relatively strong colloidal gel. $\sigma^*=1/nC$ represents the stress barrier necessary to break the gel. In the case of oscillations, $\sigma^*$ is frequency-independent. 

\section*{Acknowledgements}%
The authors thank T. Divoux and V. Grenard for useful discussions.
SM acknowledges funding from the European Research Council under the European Union's Seventh Framework Program (FP7/2007-2013) / ERC grant agreement No. 258803 and from Institut Universitaire de France. TG acknowledges funding from the Agence Nationale de la Recherche (ANR-11-PDOC-027). SBL thanks the Bernt J\"armark Foundation for Scientific Research for financial support.

Correspondence and requests for materials should be addressed to SM (sebastien.manneville@ens-lyon.fr) or TG (thomas.gibaud@ens-lyon.fr).

\footnotesize{

}
\end{document}